# Spinodal Decomposition-Enabled Halide Perovskite Double Heterostructure with Reduced Fröhlich Electron-Phonon Coupling

Yiping Wang[1], Zhizhong Chen[1], Felix Deschler[2], Xin Sun[3], Toh-Ming Lu[3], Esther Wertz[3], Jia-Mian Hu[4], Jian Shi[1, *]

[1] Department of Materials Science and Engineering, Rensselaer Polytechnic Institute, Troy, NY, 12180, USA
[2] Cavendish Laboratory, University of Cambridge, Cambridge, United Kingdom
[3] Department of Physics, Applied Physics, and Astronomy, Rensselaer Polytechnic Institute, Troy, NY, 12180, USA
[4] Department of Materials Science and Engineering, University of Wisconsin - Madison, WI, 53705, USA

[*]Correspondence: shij4@rpi.edu

**Abstract**

Epitaxial III-V semiconductor heterostructures are key components in modern microelectronics, electro-optics and optoelectronics. With superior semiconducting properties, halide perovskite materials are rising as promising candidates for coherent heterostructure devices. In this report, spinodal decomposition is proposed and experimentally implemented to produce epitaxial double heterostructures in halide perovskite system. Pristine epitaxial mixed halide perovskites rods and films were synthesized via Van der Waals epitaxy by chemical vapor deposition method. At room temperature, photon was applied as a knob to regulate the kinetics of spinodal decomposition and classic coarsening. By this approach, halide perovskite double heterostructures were created carrying epitaxial interfaces and outstanding optical properties. Reduced Fröhlich electron-phonon coupling was discovered in coherent halide double heterostructure, which is hypothetically attributed to the classic phonon confinement effect widely existing in III-V double heterostructures. The ability to develop coherent double heterostructures in halide perovskites paves an avenue to exploring halide perovskite-based quantum wells and superlattices for high-performance and low-cost optoelectronics, electro-optics and microelectronics.

The beautiful design of semiconductor heterostructures with coherent interfaces that manipulate the semiconductors' energy band landscape by breaking their translational symmetry[1] has realized a powerful control of lasing dynamics[2], charge transport behaviors[3] and electron-phonon coupling mechanisms[4] within the device. For example, with heterostructures, challenging requirements critical for electro-optics like efficient population inversion[2] modified electron-phonon coupling[4] and enhanced carrier mobility[5] have been achieved that would be otherwise impossible in a single material system. Practically over the past decades, the III-V and II-VI semiconductor families, with their favorable intrinsic properties, lattice mismatch and crystal symmetry[6-8], have always been the most popular candidates for heterostructure fabrication thanks to the advances in vapor phase epitaxy growth techniques such as Molecular Beam Epitaxy (MBE) and Metal-Organic Vapor Phase Epitaxy (MOCVD)[4, 9, 10]. Accordingly, the laboratory realization and market commercialization of heterostructure transistors[11], light-emitting diodes (LED)[12], avalanche photodetector[13], quantum well lasers[14], double heterostructure lasers[15], superlattice-based quantum cascade lasers[16] have been achieved. Recently, the discovery of new class of promising materials has added new possibilities to the heterojunction candidates. The fabrication of two-dimensional transition metal dichalcogenides coherent heterojunction[17, 18] has been under intensive investigation in the field that witnesses the unique physical properties like interlayer exciton and long-lived charge carriers[19].

A similar story is being unfolded in halide perovskites. Starting as a promising solar cell material in photovoltaic field[20], the lead halide perovskites have been later used in applications like photodetector and sensor[21, 22], LED[23] and optically pumped lasers[24, 25]. More insights have also been cast into the deep physics lying behind the superior device performance, unveiling intrinsic material properties like carrier lifetime exceeding 100 μs[26] that outperform the conventional III-V and II-VI family[4]. The success and rapid development in the perovskite field would therefore be favoring the formation of heterojunction that could give rise to properties and applications emulating those of III-V and II-VI type ones. However technically, such device is yet to be prepared probably due to the immature understanding of the material growth and the incipient instability associated with the organic-inorganic perovskite family[27, 28].

Here based on our previous experience with the growth of halide perovskite via Van der Waals (VdW) epitaxy[29] and the recent theoretical study on the understanding of the halide perovskite phase diagram[27], we showed for the first time the growth of halide perovskite double heterojunction starting from a single mixed Br-I phase followed by the photon-induced spinodal decomposition. By taking advantage of the instability of the single phase material which has long been a concern in terms of material degradation, we were able to customize the heterojunction pattern simply by controlling the photon exposure. Our device displayed excellent optical properties with distinctive photoluminescence (PL) from I-rich and Br-rich phases. The double heterojunction featured an enhanced luminescence from I-rich phase due to the carrier quenching from the band alignment and more importantly, a decreased PL linewidth was observed that can be attributed to a reduced electron-phonon coupling (Fröhlich interaction) that has been revealed typical in most successful III-V superlattices and heterostructures[4]. Our results suggest the halide perovskite family as a promising candidate for heterojunction and superlattice formation.

Fig.1 displays our design for the development and proposed properties of the halide perovskite double heterojunction. Based on our recently discovery on VdW epitaxy of halide perovskites[29], the single-phase mixed Br/I perovskite material was obtained from a Chemical Vapor Deposition (CVD) method via our recently discovered VdW epitaxy mechanism of halide perovskites by interplaying with the precursor/substrate temperature and the growth pressure as shown in Fig. 1(a). The precursor temperature plays a vital role in determining the type of halide that can be deposited. Due to the stronger bonding of Br-based perovskite[30], a higher source temperature (~320 °C) always results in the Br-based perovskite, while I-based prefers a lower one (~260 °C). The substrate temperature and growth pressure on the other hand, determine the morphology one expects to get. Generally, a lower substrate temperature and pressure would increase the nucleation rate, i.e., increasing the substrate adatom population by reducing desorption rate (by reducing temperature) and increasing flux by suppressing atom collision with lower pressure. In some extreme conditions, growth may be completely suppressed and uniform, polycrystalline film could be obtained. By increasing both temperature and pressure, with deposition rate decreased and surface diffusion encouraged, single crystalline material can be grown. A "milder" temperature and pressure usually result in thin film or rods which possesses epitaxial relationship with the mica substrate due to the Van der Waals interaction, while higher values would promote Wulff ripening[31] that overwhelms the interaction with the substrate and results in a 3D crystallite that partially satisfies Wulff construction. The inset of Fig. 1(a) shows respectively the Br-based epitaxial film, rod and 3D crystallite obtained under different growth parameters on mica substrate. A very similar result for the growth of pure I-based perovskite can be found in Fig. S1, where we have also shown the PL spectrum and PL image of the film with excitation filtered. Taking all the factors into consideration, we chose to grow the mixed halide perovskite at the source temperature of around 300 °C and a chamber pressure around 200 Torr, where some rod and square film morphologies can be expected (see Supporting information - Materials and Methods for details).

The transition from the mixed halide perovskite to distinctive I-rich and Br-rich heterojunction phases is further shown in the proposed phase diagram in Fig. 1(b) which is found qualitatively consistent with a very recent computational prediction on the phase diagram of this material system[27] (see Fig. S2). Comparatively, the general trend is followed including a spinodal decomposition regime but the phase boundary has been more asymmetric because from experimental result we believe there is limited miscibility of I in Br-rich phase compared with the situation vice versa, which will be discussed later in detail. Single phase mixed halide perovskite with I composition as $x_0$ within the spinodal regime is grown above the phase separation temperature via VdW epitaxy. During the rapid cooling process, the non-negligible VdW interaction together with the rapid cooling rate would to a great extent limit the diffusivity of I$^-$ and Br$^-$, resulting in a non-equilibrium phase at room temperature, illustrated by the shaded area in the phase diagram. Thermodynamically, upon proper activation (photon in our case), the unstable phase would be separated into two coherent I-rich and Br-rich phases, and hence comes the heterojunction. Based on this design, Fig. 1(c) depicts the atomistic model of a one-dimensional (1D) cubic mixed halide perovskite rod on the pseudo hexagonal mica (001) surface. Ideally upon laser illumination, the exposed region would be converted to the Br-rich phase with a green PL around 540 nm sandwiched by I-rich phases with red PL around 700 nm, namely a double heterojunction. The distinctive color contrast thus would be a clear signature of phase separation and the formation of heterostructure. Microscopically, a hypothesized role photon may play in the phase separation process is further illustrated in Fig. 1(d). The photon may facilitate the diffusion generally in a few aspects. First, as supported by recent simulation results about the low formation energy (e.g. 0.14 eV for reaction of $nil \rightarrow V_{MA}^{/} + V_{Pb}^{//} + 3V_I^{\bullet} + MAPbI_3$ in a Kröger-Vink representation) and migration energy (e.g. 0.58 eV for $V_I^{\bullet}$, 2.31 eV for $V_{Pb}^{//}$, 0.84 eV for $V_{MA}^{/}$) of the favored Schottky disorder in the halide perovskite material[32, 33], the photon with a much higher energy (3 eV in our case) would generate a higher vacancy concentration in the illuminated area, as shown by the schematics. This would significantly accelerate the atomic diffusion, and thereby speed up spinodal decomposition. A chemical potential-driven vacancy-assisted diffusion process is illustrated in Fig. 1(d). Secondly, several recent experimental studies report photon-induced or -assisted ion migrations phenomena in perovskites explained by a few different mechanisms[34-40]. Though these mechanisms may vary microscopically, the observed phenomenologically faster kinetics in these studies could potentially speed up spinodal decomposition process of our VdW epitaxial mixed halides system.

To evaluate the properties of halide perovskite heterostructures, we expect property modulation from two aspects: optoelectronic quantum efficiency (Fig. 1(e)) and electron-phonon coupling (Fig. 1(f)) – two features unique for semiconductor double heterostructures. Fig. 1(e) displays the proposed band alignment (Type I) of a two double heterostructures with band level information obtained from ref.[41]. It is obvious that for the I-rich phase, the conduction/valence band would serve as a sink for electrons/holes from Br-rich region. A stronger radiative recombination that corresponds to the band gap of the I-rich phase and accordingly a higher PL intensity are expected. Phonon confinement, a more unique phenomenon that applies to the double heterojunction[4], is illustrated in Fig. 1(f). Electron-Phonon interaction plays a critical role in manipulating carrier dynamics. In double heterostructures, it regulates many electro-optical properties like lasing pulsing width (formulated by Heisenberg Uncertainty principle at time and energy domain), hot carrier quenching, and carrier mobilities. At high temperature, due to the relatively high longitudinal optical (LO) phonon energy and the almost dispersionless ω-k relation, for polar crystals, LO phonons dominate the scattering process through Fröhlich interaction, in which electrons interact with the eletrostatic potential created by optical phonons[4, 42]. Namely,

$$\Phi_{LO} = \frac{F}{iq} u_{LO} \quad (1),$$

where $\Phi_{LO}$ is the induced electrostatic potential, $q$ the phonon wavevector, $u_{LO}$ the displacement of the positive ion relative to the negative ion, $F = -[4\pi N \mu \omega_{LO}^2 (\varepsilon_\infty^{-1} - \varepsilon_0^{-1})]^{1/2}$, $\mu^{-1} = M_1^{-1} + M_2^{-1}$, $N$ number of unit cells per unit volume of the crystal, $\mu$ reduced mass of the primitive cell, $M_1$ and $M_2$ masses of the two atoms inside the primitive cell, respectively, $\omega_{LO}$ LO phonon frequency, $\varepsilon_\infty$ and $\varepsilon_o$ the high- and low-frequency dielectric constants, respectively. The Hamiltonian $H_{FR}$ for the electron-phonon coupling would be directly linked with the potential by:

$$H_{FR} = \sum_q -e\,\Phi_{LO} = \sum_q \frac{ieF}{q} u_{LO} \quad (2),$$

where $e$ is the electron charge. The ability to manipulate Fröhlich interaction by phonon confinement in double heterojunction clears the path for modifying electronic transport property and recombination dynamics. According to a recent Raman study on different halides[30] (see further discussion on Fig. S3) and the fact of different bonding strengths of different halides, a large gap between the well (e.g. Br-rich) and barrier (e.g. I-rich) of a halide double heterostructure in the LO phonon (a dominant Fröhlich interaction LO phonon would likely be Pb-X stretching mode, see Fig. S3) vibrational frequency (simplified by springs of different colors in Fig. 1(f)) would inevitably induce a strong phonon confinement effect. The confined phonons, for example, in a Br-rich phase well between two I-rich phase barriers would have quantized phonon wavevector of

$$q_m = \frac{m\pi}{L} \quad (3),$$

where $L$ is the width of the Br-rich phase well and $m$ an integer. Under phonon confinement approximation, both mechanical and Maxwell's boundary conditions need to be satisfied. The lower part of Fig. 1(f) shows the schematic potential profiles (Macroscopic Huang-Zhu model[43] is applied for qualitative sketching here as it deals with both boundary conditions) for two double heterojunctions with different widths of Br-rich phase well under two exampled quantized phonon wavevectors. Clearly, a long wavelength phonon (small $q$) contributes more to the induced potential. If the phonon is confined to a larger $q$ number by a reduced width $L$, the potential is greatly reduced and Fröhlich interaction suppressed. The $1/q$ dependence indicates that the wavevectors with long wavelength are more dominant in Fröhlich interaction. Therefore, we expect to observe such modification of Fröhlich coupling in a halide perovskite double heterostructure. Over decades, such confinement has already been well achieved in III-V heterostructures and superlattices[44-46].

The growth and characterization results guided by the design in Fig. 1 are further shown from Fig. 2. Fig. 2(a) and (b) display the optical images of 1D epitaxial mixed halide perovskite rods obtained with the growth parameters mentioned above. It can be seen that the rods align themselves either with 60° or 120° angle, which is exactly the pseudo six fold symmetry of the (001) mica surface. Thicker rods and crystallites can be found on substrates closer to the precursor where the temperature is higher (Fig. S4). The most interesting feature of the mixed halide perovskite rod is the dramatic change upon laser illumination as shown in Fig. 2(c) and (d). The rod showed a dim orange-like PL upon instant illumination (not shown here) which turns into very bright reddish orange (Fig. 2(c)) within a few seconds that lights up the two ends due to waveguide effect. Within several minutes, the area exposed to the laser turned into bright green PL, but the tips of the rod still preserves red color. The rapid change of PL in mixed I/Br perovskite has also been observed elsewhere in spin-coated film[39]. To better understand the complicated time-dependent PL change of the mixed halide perovskite rod, a series of time-lapsed optical images were taken as shown in Fig. 2(e-h) with intervals of 1 s. With this we were able to capture the dim PL as shown in Fig. 2(e). Within 2 s, the orange like PL color rapidly converts to bright red color in Fig. 2(f), which later undergoes a gradual change to green color within 80 s Fig. 2(g-h). Another set of time dependent images with similar proceedings can be found in Fig. S5. We attribute this change of PL to a light induced phase separation of mixed halide perovskite into Br-rich and I-rich phases that would be studied in detail later. By taking advantage of such phenomenon, we were able to pattern the rod simply by controlling the laser exposure with designed intervals into a 1D pseudo-superlattice, as shown in Fig. 2(i). The illuminated and unilluminated regions present themselves with different contrast under optical microscope that are easily recognizable. Upon weak uniform laser illumination, the I-rich and Br-rich phases would show completely different PL colors that confirm the possible formation of 1D lateral double heterostructures as shown in Fig. 2(j). To confirm this and better understand the structural information of the heterojunction, TEM electron diffraction analysis was carried out. It was observed that the I rich phase is quite vulnerable under the high tension and would quickly decompose but the Br rich phase turns out to be much more stable. Fig. 2(k) shows the Selected Area Electron Diffraction (SAED) pattern at pure Br-rich area. From the image a clear square pattern that falls into the cubic symmetry can be observed. We indexed the diffraction spots to be (220) and ($2\bar{2}0$) of the Br-rich phase with a lattice constant of $a_1 = 6.0$ Å. The value is slighter large than that of pure $MAPbBr_3$ (5.93 Å)[47], which is reasonable due to the slight inclusion of I element. The SAED pattern at the interface is further shown in Fig. 2(l). Clearly, apart from the square pattern from the Br-rich phase, we observed two more sets of diffraction spots with same symmetry but larger and different *d*-spacing marked by red ($a_2 = 6.30$ Å which is slightly smaller than pure $MAPbI_3$[48]) and green ($a_3 = 6.19$ Å which is between pure $MAPbBr_3$ and $MAPbI_3$) dashed circles. The alignment of three sets of diffractions indicates their crystals form well epitaxial relations. The diffraction patterns with exaggerated contrast can be found in Fig. S6. We interpret this as a strong evidence of spinodal decomposition where Br-rich (marked by the blue circles), I-rich (red circles) and residual parent phases (green circles) coexist. This type of SAED patterns have been well documented in literature on explaining spinodal decomposition in III-V systems[49]. To better illustrate the spinodal structure in real space, we present a tentative schematics of the lattice model in Fig. 2(m) based on the diffraction pattern and PL images. The Br-rich phase with a smaller lattice constant surrounded by I-rich and residual parent phases with different and larger lattice together forms epitaxial interfaces.

Along with structural evolution, in a spinodal decomposition process, we anticipate composition transition as well. A composition transition indicates a transient process on the optical properties of the materials. Therefore, to confirm and understand more quantitatively the laser-induced phase separation of the mixed halide perovskite, we obtained the PL spectra at different stages of laser illumination as shown in Fig. 3(a)-(b), which correspond to respectively the green and red regime. It should be noted that the spectra from Fig. 3(a) were collected separately from Fig. 3(b) due to the limitation of wavelength range of our CCD therefore spectra continuity is not expected between Fig. 3(a) and (b) due to the stochastic nature of the kinetic process. While the initial "dim PL" stage is too fast for spectrum acquisition, we were able to capture the remaining stages as shown by the four curves. A schematic is shown in Fig. 3(c) to help explain the details of the phase change deduced from the PL spectra. Upon laser illumination, the dim PL image shown in Fig. 2(e) is supposed to represent the spectrum of the single phase mixed halide but within very short time, the light could facilitate spinodal decomposition that gives rise to the blue curve in Fig. 3(a) and (b). As a typical feature of spinodal decomposition, the phase separation is supposed to happen instantaneously across the illuminated region without any nucleation barrier. Very small domains of I-rich and Br-rich phases are supposed to form by gradually changing their halide composition from the mixed to the thermodynamically preferred I-rich and Br-rich one (The first stage in Fig. 3(c)). Therefore from the PL change, the transition from Fig. 2(e) to (f) would proceed very fast and also the spectrum would present itself a very broad peak (as a result of spatially varied spinodal decomposition rate due to the non-uniform photon illumination doses out of a Gaussian laser beam) that covers almost the whole green-red region from 520 to 750 nm. This is exactly what we have observed in the blue curve of Fig. 3(a) and (b). Due to the large lattice mismatch (~ 5% deduced from our SAED and ~ 8% from literature[48]) between the I- and Br- perovskite and thus possibly giant strain energy in an epitaxial junction, the interface area would tend to become minimized. This is very common in plenty of alloys[50-52]. Thus accordingly, at the later stage of spinodal decomposition[50-55], the small domains are likely to "merge" (i.e. *coarsening*) to reach a more stable configuration, as shown in the second step of Fig. 3(c). During such process, with the fast diffusion of halide ions (shorter than a few tens of seconds in our case), transient phases would appear[52]. Due to the high mobility of halide ions, these transient phases would carry comparative Br and I concentration[52] leading to transient PL spectra featuring very unstable mixed halide phases. Since the I ion is more mobile than the Br one[30], the Br-rich phase would prefer to remain in the illuminated region while the I-rich phase would migrate away. From the PL spectra, such process would show as a blue shift of the peak within the red regime (the formation of the transient phase) and the increase of peak intensity in the green regime (the tendency for the Br-rich phase to remain under the laser spot). Such trend is observed and reflected by the green and yellow PL spectra in Fig. 3(a) and (b). In Fig. 3(b), the I-rich PL peak evolves following the time sequence marked as "1→2→3→4", in which peak shift corresponding to "→2→3" serves as a strong evidence on the proposed coarsening process. As a further evidence, at this stage (i.e. "→2→3"), by tuning the laser illumination to be very weak but uniform (see Supporting information – Materials and Methods), we were able to observe a gradual change of PL color from green to red from the originally laser illumination center to the far edge, as displayed in the left inset of Fig. 3(c). This suggests the existence of some transient phases. After domains become large, coarsening would approach ceasing asymptotically. The transient region should become minimized, which is indeed observed as shown in the right inset of Fig. 3(c). Accordingly, the PL spectrum is expected to show two distinctive peaks corresponding to the two thermodynamically stable phases. At the stage of "→4" of Fig. 3(a) and (b), the hypothesis is partly confirmed by the observation of a single sharp peak around 520 nm in the green regime and a red shifted but broad peak around 670 nm in the red regime. We believe the broadening of the red shifted peak may be due to the spinodal decomposition in previously unilluminated regime, which may be initialized and facilitated by green PL of the Br-rich phase and the waveguide role of the rod itself. In addition, the photo recycling property of halide perovskite recently discovered[56] may encourage the process as well. Interestingly, we found that the unpatterned rod exposed to room light for two weeks shows a gradual transition of PL colors upon uniform and weak excitation illumination as displayed in Fig. S7. The beautiful blend of colors is another evidence for light-induced spinodal decomposition and coarsening process. The weak room light perhaps greatly reduces the kinetics and therefore appears the more long living transient phases.

Apart from the photon-driven spinodal and coarsening hypothesis, we investigated the possible heating effect. Fig. S8 shows the optical image of a single rod at 100 °C, with very bright red PL visible. Phase separation and coarsening still proceed, but with a faster rate, which is reasonable due to the temperature promoted diffusivity. However, it is found that temperature alone could not trigger appreciable spinodal decomposition and coarsening (See Fig. S8). Therefore, the heating would only be secondary compared with the photon effect. Energy Dispersive X-ray spectroscopy (EDX) was done as the elemental composition analysis to support our hypothesis as shown in Fig. 3(d). Unlike the optical images, the Scanning Electron Microscope (SEM) image of the patterned rod showed very small contrast. The contrast which is minor in SEM but obvious in optical microscope is a good evidence that the rods were not likely to experience significant materials loss (e.g. chemical decomposition) or morphology evolution during photon illumination, another evidence on spinodal decomposition and coarsening process. The resultant EDX spectra for both phases clearly show a higher peak for I/Br in the I-rich/Br-rich phase (highlighted by the two dashed circles), respectively, though the e-beam exposure would quickly damage the I-rich phase. Other elements including K, Al, Si and O that correspond to the mica substrate and Pb for the rod showed almost no difference in either region, which validates our previous comparison on I/Br content. We further confirmed the double heterojunction by carrying out a PL mapping shown in Fig. 3(e-f). By mapping the PL respectively at the green (540 nm) and red regime (710 nm), we were able to reconstruct the Br-rich and I-rich phases. Another important information obtained from the PL mapping is that the I-rich phase always display higher PL intensity at the interface, consistent with our hypothesis in Fig. 1(e). Raman characterization was also done to check the vibrational difference in two phases to confirm the heterojunction. Fig. S9 shows the Raman spectra for both phases with clearly different peak positions. To be more cautious, we compared the spectrum with that of $PbI_2$, a most probable chemical decomposition product and found a large mismatch, suggesting the absence of materials loss. In Fig. S9, we also carried out Raman mapping which reveals the heterostructure in a similar way as PL mapping. To further confirm our hypothesis on the kinetic process in mixed halides, we also performed phase-field simulations (see Supporting information – Materials and Methods for details) to qualitatively illustrate the spinodal decomposition (from Fig. 3(g) to Fig. 3(h)), and subsequent coarsening, that is, two Br-rich sections (black contrast) (Fig. 3(h)) merge into a single Br-rich section (Fig. 3(i)). These simulated results are found consistent with experimental observations.

In addition to these 1D rods, similar photon illumination-enabled kinetics has been found in 2D mixed halide perovskite flake as well. The corresponding results are shown in Fig. S10 including pristine growth results, Atomic Force Microscope (AFM) characterizations, PL optical images, PL spectra upon laser illumination and PL mapping. These results show that with adequate control of the photon doses, lateral 2D heterojunction with desired features could be achieved. More detailed discussion can be found in the supplementary information.

As reduced Fröhlich electron-phonon coupling is a unique feature of double heterostructures compared to their homogeneous structures, we carried spectroscopy study on revealing the magnitude of Fröhlich coupling in our halide perovskite heterostructures (Fig. 4). A significant and quantifiable electron-phonon coupling effect is the broadening of PL peaks, which was employed by previous studies to understand the electron-phonon interaction in traditional III-V semiconductors[4] and even recently in single phase perovskite[57, 58]. Close to room temperature, it is found that dominant electron-phonon coupling in single phase halide perovskite is Fröhlich coupling[58], consistent with most III-V semiconductors[4]. Fig. 4(a) shows the room temperature PL spectra of the Br-rich phase in the double heterojunction together with those of homogeneous structure (single phase Br-based perovskite film or rod by VdW epitaxy) and bulk single crystal phase by solution approach. Apart from the PL peak position, we notice a significant decrease of PL peak Full Width at Half Maximum (FWHM) for the double heterojunction case. Such phenomenon is observed in both the 1D and 2D double heterojunctions. We attempt to attribute it to the reduced Fröhlich coupling due to phonon confinement, which is a typical consequence of double heterostructures in III-V systems as discussed earlier. To confirm this hypothesis, we performed temperature dependent PL study of all three kinds of samples. As an example, Fig. 4(b) shows the PL spectra of homogeneous structure ranging from -50 °C to 70 °C. Fig. S11 presents the temperature-dependent PL for Br-based phase in the form of both double heterostructure and bulk single crystal. From the stack of spectra in Fig. 4(b) we can observe a general blue shift of PL peak with increasing temperature, which is consistent with the abnormal temperature-dependent band gap of halide perovskite[57]. A red shift of peak from -50 °C to -40 °C is observed that could be assigned to the structural phase change from cubic to tetragonal of the perovskite material. Fig. 4(c) plots the change of band gap over temperature for both homogeneous film structure and double heterostructure and we see apart from a general blue shift of peak position for the homogeneous epitaxial film, the band gap of both samples follows a very similar abnormal temperature dependence trend, which further confirms the material. The FWHM, on the other hand, showed a quite different story as shown in Fig. 4(d). We witness a general decrease of FWHM for the double heterojunction case as in Fig. 4(a) over the whole temperature range. More importantly, the FWHW for the homogeneous phase decreases gradually with decreasing temperature while for the double heterojunction temperature seems to be less perturbing to the FWHM. From this and our discussion earlier we believe it is where the electron - LO phonon interaction (Fröhlich interaction) plays a vital role. The temperature dependence of the linewidth of the PL is formulated as follows[58]:

$$\Gamma(T) = \Gamma_0 + \Gamma_{acoustic} + \Gamma_{LO} + \Gamma_{impurity} = \Gamma_0 + \gamma_{acoustic}T + \gamma_{LO}\frac{1}{(e^{\frac{E_{LO}}{k_BT}}-1)} + \gamma_{impurity}e^{-e^{\frac{E_b}{k_BT}}} \quad (4),$$

where $\Gamma$ is linewidth, $T$ temperature, $\Gamma_0$ temperature-indepdendent broadening term due to disorder, $\Gamma_{acoustic}$ linewidth broadening term associated with acoustic phonon scattering, $\Gamma_{LO}$ linewidth broadening term associated with Fröhlich phonon scattering, $\Gamma_{impurity}$ broadening term from scattering associated with ionized impurities, $E_b$ binding energy, $\gamma_{acoustic}$, $\gamma_{LO}$, $\gamma_{impurity}$ charge carrier-phonon coupling strengths associated with acoustic phonon, LO phonon and impurity, respectively, $E_{LO}$ LO phonon energy which is equal to $\hbar\omega_{LO}$, $\omega_{LO}$ polar LO phonon frequency, $k_B$ Boltzman constant. According to Wright et al[58], at above 150 K, in halide perovskites, both acoustic phonon scattering and ionized impurities scattering are negligible. Therefore, a combination of $\Gamma_0$ and $\Gamma_{LO}$ can well describe the overall electron-phonon coupling of both homogeneous film and double heterostructure at the temperature window applied in our study. As $\Gamma_0$ (temperature independent) and $\Gamma_{LO}$ (Bose-Einstein statistics) carry completely different temperature dependency, fitting of PL linewidth would help uncover the coupling strengths. The fitting results for both homogeneous phase and double heterostructure are displayed in Fig. 4(d). The value of $\hbar\omega_{LO}$ is obtained from ref. [58]. Both qualitatively from the graphs and quantitatively from the fitting results, for the double heterojunction, the $\gamma_{LO}$ value is significantly reduced compared to homogeneous film and the dominant broadening factor becomes the intrinsic parameter $\Gamma_0$. This could likely be attributed to phonon confinement effect in double heterostructures discussed in Fig. 1 part. The reduced Fröhlich coupling is nearly absent in homogeneous phase (when compared to bulk single crystal as shown in Fig. 4(a)) probably due to lack of phonon mismatch. Therefore the halide perovskite double heterojunction in the perspective of electron-phonon coupling is very promising and unique for electro-optics, optoelectronics and microelectronics.

In conclusion, by taking advantage of the spinodal decomposition, we have for the first time showed the successful growth of the halide perovskite double heterojunction by advancing kinetics via photons. We proved the possibility to customize the heterojunction pattern while at the same time preserving good optical properties. We have suggested a phenomenological mechanism and proceedings of the spinodal decomposition by structure analysis and spectroscopy study. With double heterostructures, we have observed a reduction in Fröhlich electron-phonon coupling likely due to the phonon confinement effect. As a step forward, the spinodal decomposition could possibly be further utilized for finer heterostructures and even superlattices with wavelength at a few nanometers scale, which have already been proved successful in conventional III-V family[59, 60]. Our work could potentially introduce a new set of technically useful, high-performance optoelectronic, electro-optic, microelectronic materials and devices in the forms of heterostructures and superlattices.

**Supporting Information** is available online.

**Acknowledgments**
Financial support was provided by the US National Science Foundation under Grant No. CMMI 1550941. Jian Shi acknowledges the helpful discussion with Xiaodong Xu in developing the project.

**Authors Contribution**
Y.W. and J.S. conceived the study. Y.W. grew the materials, carried PL and Raman spectroscopy study, conducted TEM and SAED characterizations and AFM imaging. Y.W. and Z.C. performed SEM and EDX study. J.H. carried phase field simulation. Y.W. and J.S. analyzed the data and wrote the manuscript. All authors discussed the results and edited the manuscript.

**Competing Financial Interests**
The authors declare no competing financial interests.

# Figures and figure captions

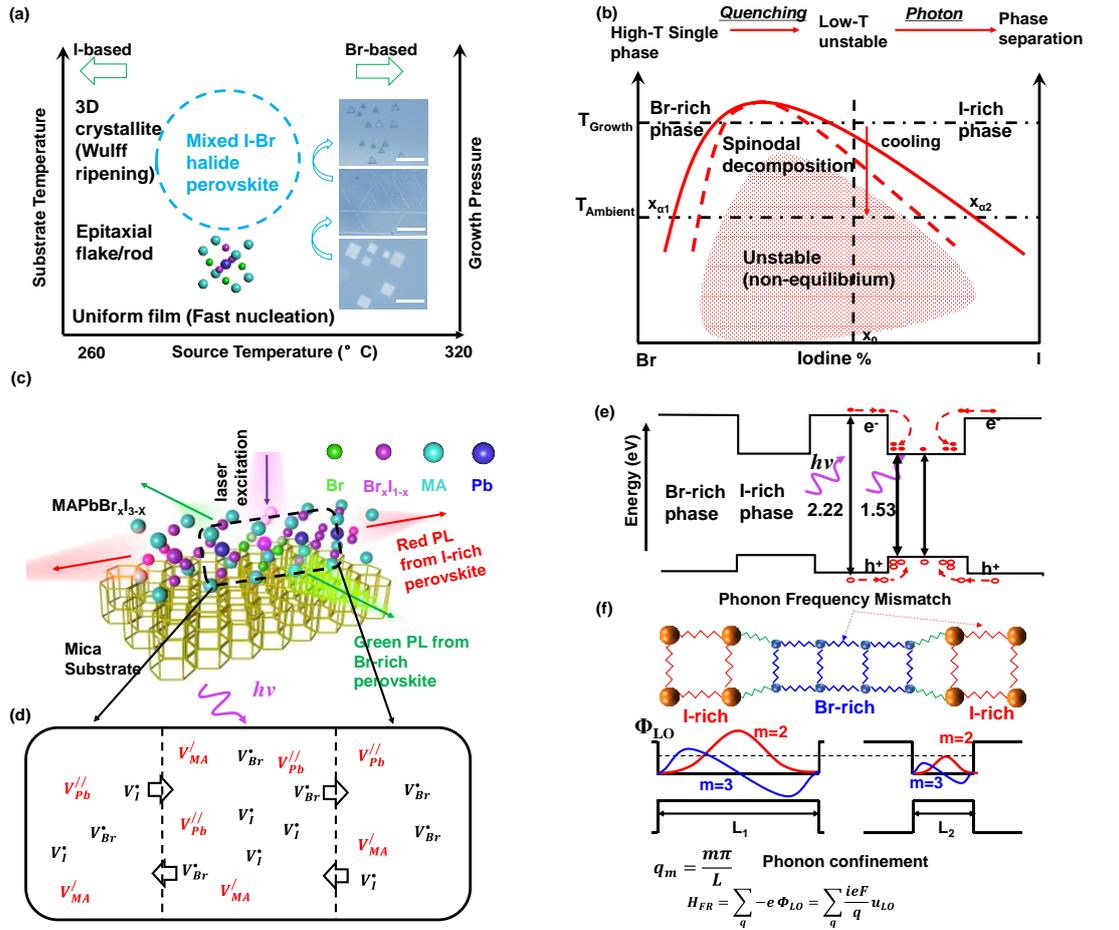

**Fig. 1 Design and proposed properties of the halide perovskite double heterojunction** (a) Schematic diagram showing the effect of source/substrate temperature and chamber pressure on the obtained type and morphologies of halide perovskite. The region where desired mixed halide perovskite is grown can thus be located. The inset shows the typical growth result of different morphologies of the pure Br perovskite. (scale bar: (up to bottom) 30 μm; 50 μm; 20 μm). (b) Schematic phase diagram of the I/Br mixed halide perovskite. The as-grown single phase material with I composition of $x_0$ would go through a spinodal decomposition into coherent $\alpha_1$ and $\alpha_2$ phases facilitated by the photon. (c) Atomic model of the cubic mixed halide perovskite heterojunction on hexagonal mica substrate. Ideally, the Br-rich and I-rich phase would give out completely different PL colors upon laser excitation once the phase separation is complete. (d) Schematics of the photon promoted diffusion in microscopic scale under photon illumination. Kröger–Vink notation is applied to index defects. (e) Diagram of electron and hole quenching at the I-rich phase resulted from the Type-I band alignment of the two perovskite phases. (f) Schematics of the phonon mismatch (upper part) giving rise to the phonon confinement. Diagrams of the induced Fröhlich potential (lower part, qualitatively sketched based on Huang-Zhu model) for different scale of double heterojunctions under two quantized phonon wavevectors (m = 2, 3). The equations at the bottom show the quantized phonon wavevectors and the Fröhlich electron-phonon interaction Hamiltonian (see more description in main text).

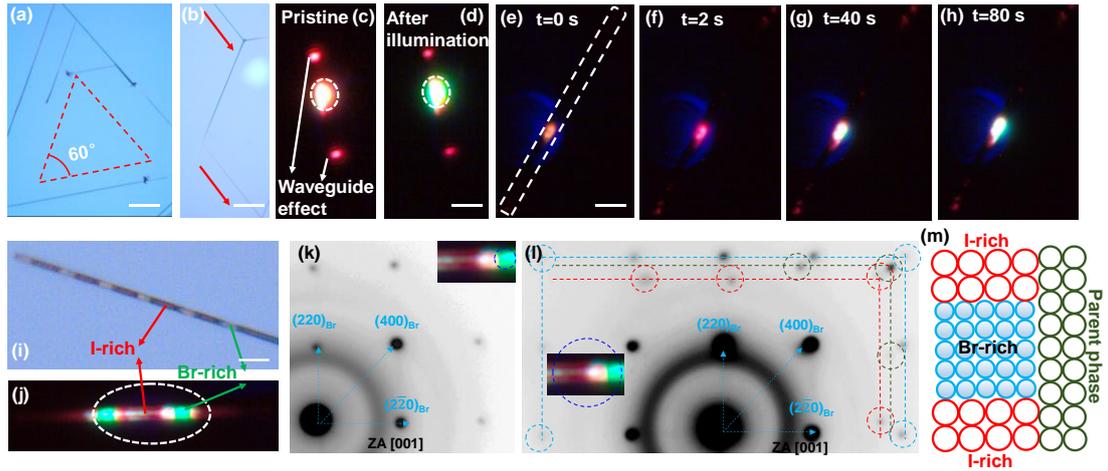

**Fig. 2 Experimental results for the 1D mixed halide perovskite double heterojunction.** (a-b) Optical images of the mixed halide perovskite epitaxial rod grown on mica substrate. The rods are either 60° (a) or 120° (b) aligned which follows the symmetry of mica. (c-d) PL images of the 1D rod with large difference before and right after laser illumination (c) and minute later (d). The lightening at the ends of the rod is a result from the waveguide effect. (e-h) Time lapsed PL images of one rod from t=0 s (e) to t=80 s (h). Significant PL changes in color and intensity can be observed. When overall intensity is high, the waveguide effect is visible. The outline of the rod is highlighted by dashed white lines. (i) Optical image of multiple double heterojunctions patterned on a single rod by manipulating the laser location. The two difference phases show clear contrast. (j) PL image of the double heterojunctions with green color from Br-rich phase and red color from I-rich phase upon uniform laser illumination. (k) SAED pattern of the Br-rich phase, collected from the region marked in the inset image. Square patterns with [001] as the zone axis can be indexed. (l) SAED pattern at the interface. Three sets of aligned patterns marked by red, green and blue dashed circles can be observed. (m) Proposed lattice model based on SAED patterns and PL images. (scale bar: (a) 50 μm; (b) 160 μm; (c-d) 4 μm; (e-h) 5 μm; (i) 4 μm)

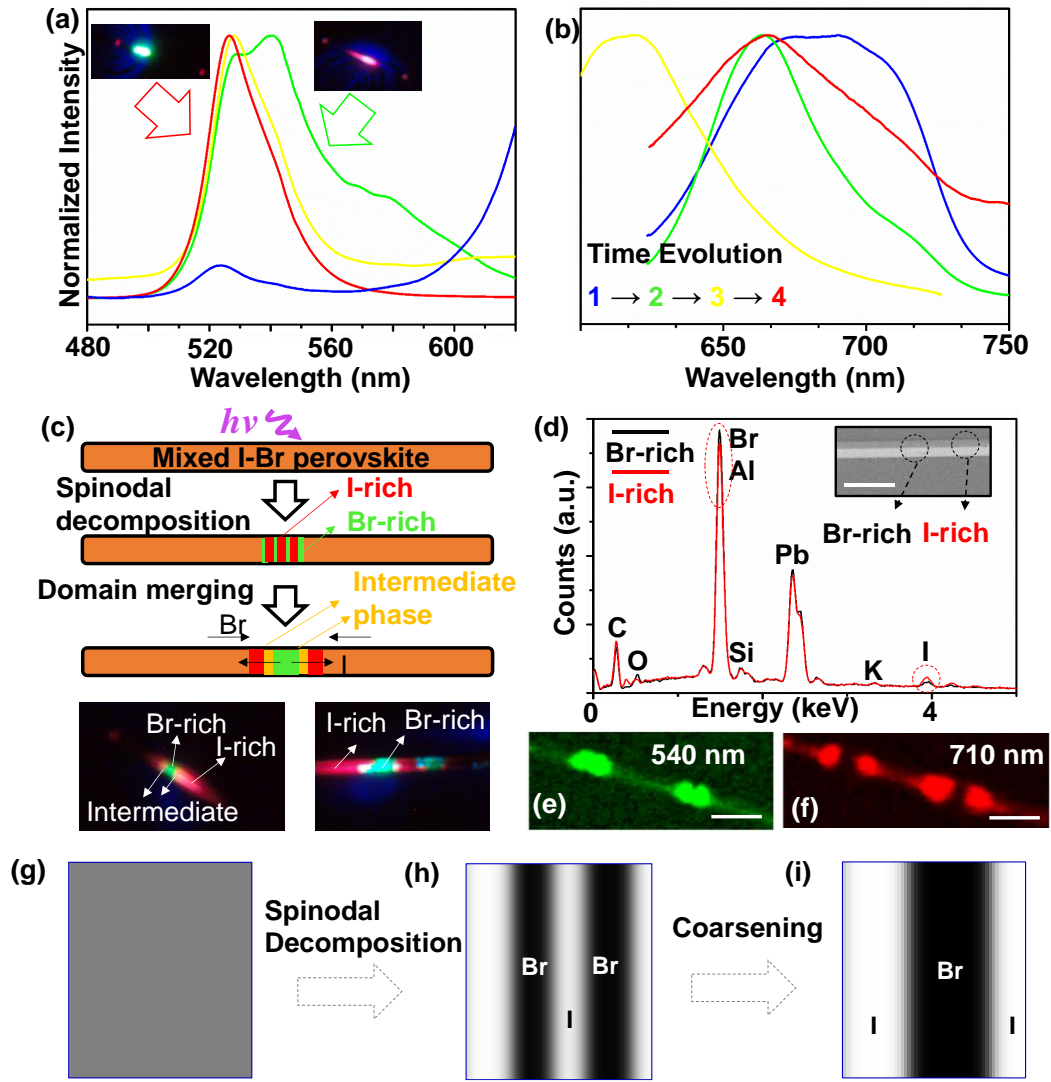

**Fig. 3 Spectroscopy and elemental analysis of the double heterojunction.** (a-b) Time dependent PL spectra of the halide perovskite rod during laser illumination process in the green color regime (a) and red color regime (b). The inset of (a) shows the PL images of the rod at the second and fourth stage, respectively. The color of the curves indicate time evolution as shown in (b). (c) Schematics showing the proposed microscopic process during illumination deduced from (a) and (b). The overall process include the nucleation-barrier free spinodal decomposition and classic coarsening. The two lower insets show the halfway stage where the intermediate phase displays transient orange PL (left) and the fourth stage where distinctive color difference can be observed (right) (white color indicates PL intensity saturation). (d) EDX analysis of the double heterojunction. The EDX spectra for the two phases show clear I and Br content difference. The elements (K, Al, Si, O) in the mica substrate and Pb are consistent at two locations. The inset shows the SEM image of the heterojunction with little contrast between the two phases. (e-f) PL mapping of the double heterojunction at 540 nm (e) and 710 nm (f). The counts from I-rich phases at 540 nm come from some Raman signals since a 532 nm excitation is used. (g)-(i) Phase-field simulation of spinodal decomposition and coarsening in perovskites. (scale bar: (d) 4 μm; (e-f) 2 μm)

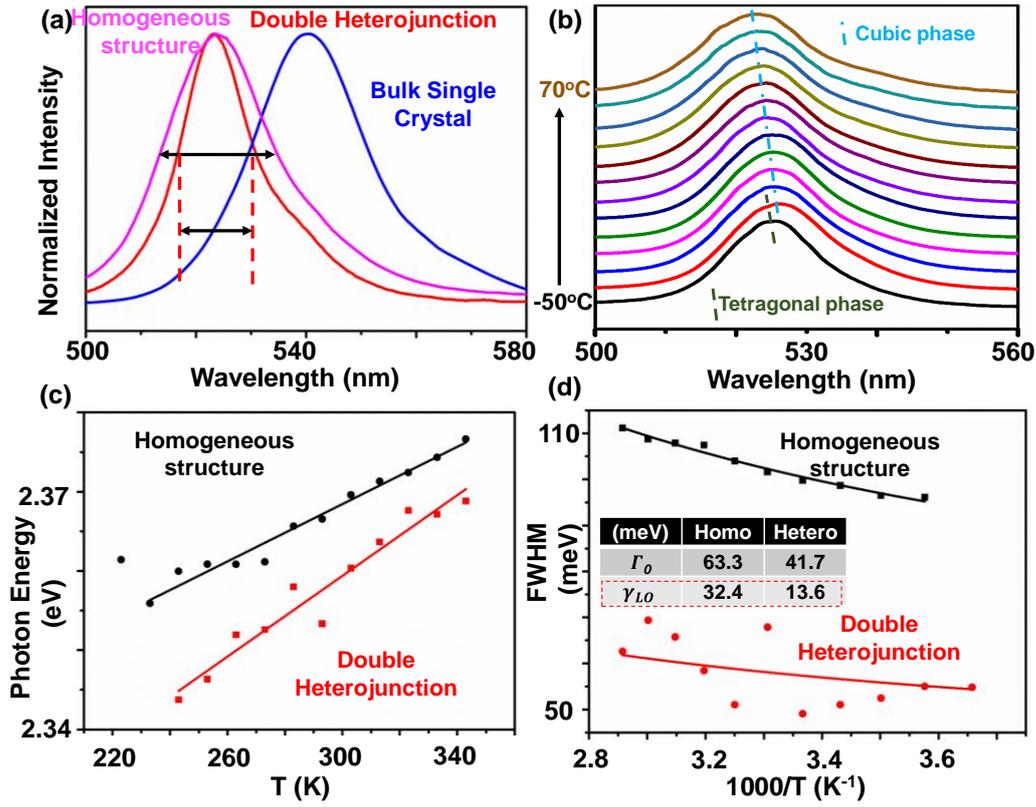

**Fig. 4 Spectroscopy study of the electron-phonon coupling in double heterojunction.** (a) PL spectra of the Br-based homogeneous structure (purple), bulk single crystal (blue) and double heterojunction (red). A significant reduction in the PL linewidth of double heterostructure can be observed. (b) Temperature dependent PL spectra of homogeneous film structure from -50 °C to 70 °C. Besides the abnormal temperature dependent peak shift, a transition from cubic (blue dashed line) to tetragonal phase (black dashed line) at -50 °C is also observed. (c) Summary of the band gap-temperature plots of both homogeneous structure (black) and double heterojunction (red). (d) FWHM-temperature plots of both homogeneous structure (black) and double heterojunction (red). Insets display the fitting parameters.

# Supplementary Information

**Spinodal Decomposition-Enabled Halide Perovskite Double Heterostructure with Reduced Fröhlich Electron-Phonon Coupling**

Yiping Wang[1], Zhizhong Chen[1], Felix Deschler[2], Xin Sun[3], Toh-Ming Lu[3], Esther Wertz[3], Jia-Mian Hu[4], Jian Shi[1, *]

[1] Department of Materials Science and Engineering, Rensselaer Polytechnic Institute, Troy, NY, 12180, USA
[2] Cavendish Laboratory, University of Cambridge, Cambridge, United Kingdom
[3] Department of Physics, Applied Physics, and Astronomy, Rensselaer Polytechnic Institute, Troy, NY, 12180, USA
[4] Department of Materials Science and Engineering, University of Wisconsin - Madison, Madison WI, 53705, USA

*Correspondence: shij4@rpi.edu

**Contents:**
Materials and Methods
Supplementary Figures S1 – S11
References

# Materials and Methods I

**Chemical vapor deposition (CVD)/Van der Waals epitaxial synthesis of single phase mixed I/Br perovskite**

Mixture of powdered Lead(II) Bromide($PbBr_2$, 99%, Sigma-Aldrich) and Lead(II) Iodide($PbI_2$, 99%, Sigma-Aldrich) was placed in the furnace heating center with the heating temperature controlled at around 300 °C, while the mixture of MAI and MABr was placed about 7 cm away from the lead halide in the upper stream due to a lower melting point. The MAX powder was prepared according to a previous study[1]. Fresh cleaved muscovite mica substrates (SPI Grade V-5) with (001) face exposed were placed in the downstream. Prior to deposition, the base pressure of the system was pumped to 0.5 Torr after which a 30 sccm of Argon was flowed to maintain the pressure at 170 Torr before deposition. The chamber temperature rose from room temperature to the deposition temperature rapidly in 5 min. The deposition process lasted for 20 minutes before finally the furnace was shut down. The furnace was cooled down to around 90 °C before the mica substrates were taken out. The mixed perovskite epitaxial rod and film were found in most cases on the substrate about 4-6 cm from the lead halide precursor in the downstream part.

The growth for pure Br and I based perovskite followed similar procedures but at different source temperature, i.e., around 320° C for the Br-based and 260° C for the I-based.

# Materials and Methods II

**Microscopy Characterization**
Morphology of the halide perovskite heterojunction was characterized by a Nikon Eclipse Ti-S inverted optical microscope. Transmission Electron Microscope JEOL JEM-2010 was used to characterize the crystallographic information of the different perovskite phases. Scanning Electron Microscope ZEISS 1540 EsB was used for characterizing the rod morphology and performing the Energy Dispersive X-ray compositional analysis.

**Raman Spectroscopy Characterization, Raman and PL mapping**
Vibrational modes of the the halide perovskite heterojunction were obtained from Raman spectrum collected using a Witec Alpha 300 confocal Raman microscope with an excitation source of cw 532 nm under magnification of 100×. The Raman and PL mapping were done under the same confocal setup with an integration time of 0.1 s per point.

**Photoluminescence Characterization**
The PL characterization was done via a customized PL system consisting of a Picoquant 405 nm pulsed laser with a 2 mW power, the same optical microscope that focuses the laser via a 50× objective lens, a Princeton Instruments SP-2358 spectrograph and a Thorlabs 4 Megapixel Monochrome Scientific CCD Camera. The uniform laser illumination was obtained by manipulating a *f*=500 mm convex length before the optical microscope to focus the laser at the focus point of the objective lens in the microscope.

**Temperature dependent PL characterization**
Temperature dependent PL was carried out by hooking the sample in a INSTEC HCS302 microscope cryostat which can tune the temperature by a MK2000 temperature controller. The cryostat was put under the microscope to get the PL spectrum at each temperature.

**Atomic Force Microscope Characterization**
Multimode™ Atomic Force Microscope is used to obtain the film thickness information using both contact and tapping AFM mode.

**Phase-Field Simulation**
Within the framework of phase-field diffuse interface theory[2], we computationally model the spinodal decomposition and the subsequent coarsening by solving Cahn-Hilliard equation[3], that is,

$$\frac{\partial C_{Br}}{\partial t} = \nabla M \nabla \left[ \frac{\partial f_0(C_{Br})}{\partial C_{Br}} - \kappa \nabla^2 C_{Br} \right] \quad (S1)$$

where the composition variable $C_{Br}$ is utilized to represent the Br fraction in Br-rich phase (for simplicity and due to lack of experimental data, we approximate $C_{Br} = 1$ which does not lose generality), I-rich phase ($C_{Br} = 0$ for a same reason as Br-rich phase), and their interface ($0 < C_{Br} < 1$); the kinetic coefficient $M = C_{Br}C_I(C_{Br}M_I + C_I M_{Br})$, where $M_{Br}$ and $M_I$ are the mobility of the two main diffusing species in the two constituent phases (Br and I herein, respectively), and $C_I = 1 - C_{Br}$; the gradient energy coefficient $\kappa$ (= 4) is related to the interface energy; $f_0$ is the local chemical free energy density, which, for illustration, is assumed to obey a double well function, i.e., $f_0 = C_{Br}^2(1-C_{Br})^2$. To simulate the rod-type structure in experiment, we consider for simplicity a one-dimensional discretized system of 256 grids with a periodic boundary condition. Equation S1 is then solved using a semi-implicit Fourier spectral method[4]. Note that the images in Fig. 3(g)-(i) show the morphological evolution in the same local region of the simulation zone that was found to agree qualitatively with experimental observations.

# Supplementary Figures

# Fig. S1

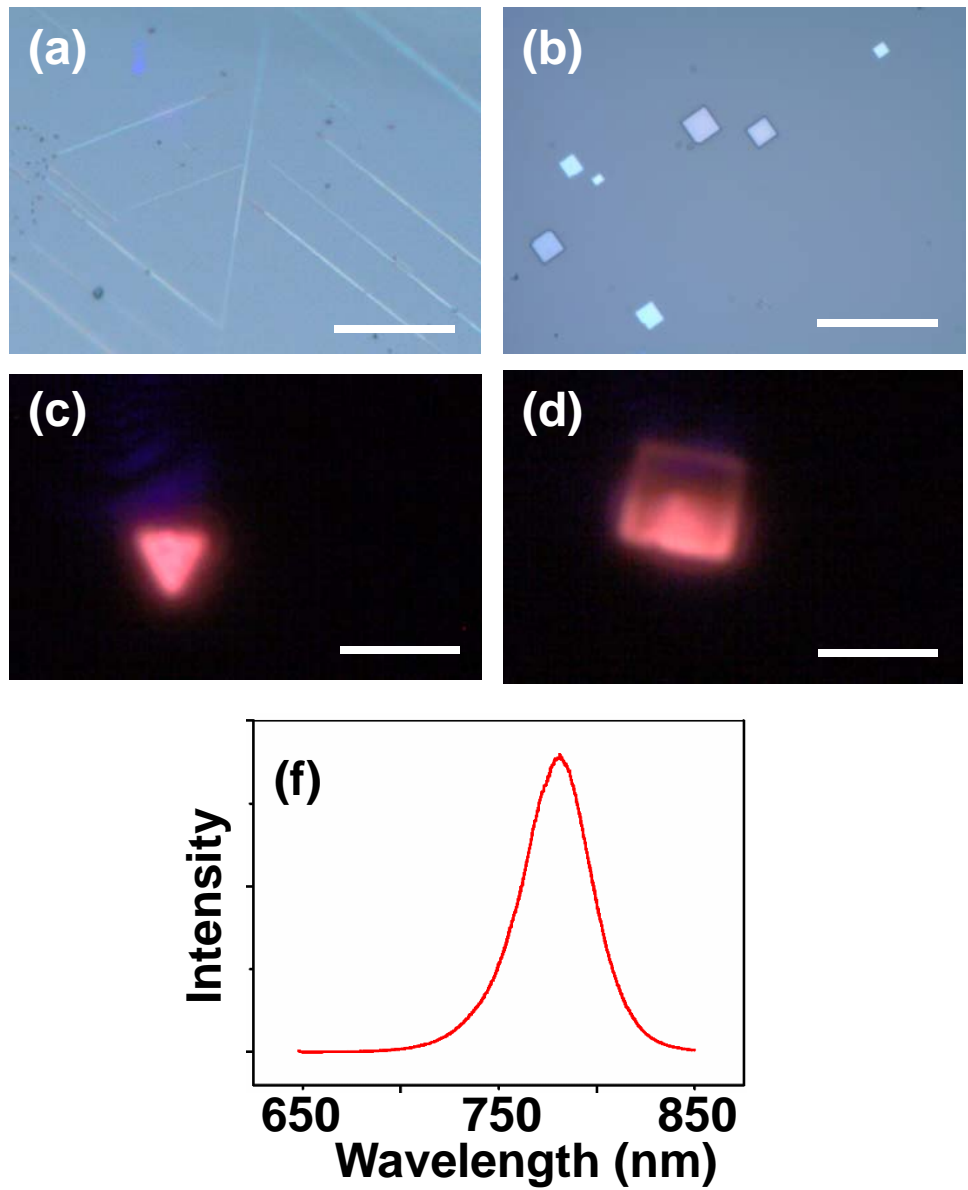

Fig. S1 Optical and PL images and PL spectrum of pure I-based perovskite. (a) epitaxial rod of MAPbI$_3$. (b) Epitaxial square flakes of MAPbI$_3$, the square can be observed to align in one direction. (c-d) PL images of a bulk pyramid crystallite (c) and square thin flake (d). The bright red color correspond to the band gap of the I-based perovskite. (f) PL spectrum of the I-based perovskite with a sharp peak at around 780 nm, indicating good crystallinity. (scale bar: (a) 10 µm; (b) 100 µm; (c-d) 5 µm; )

**Fig. S2**

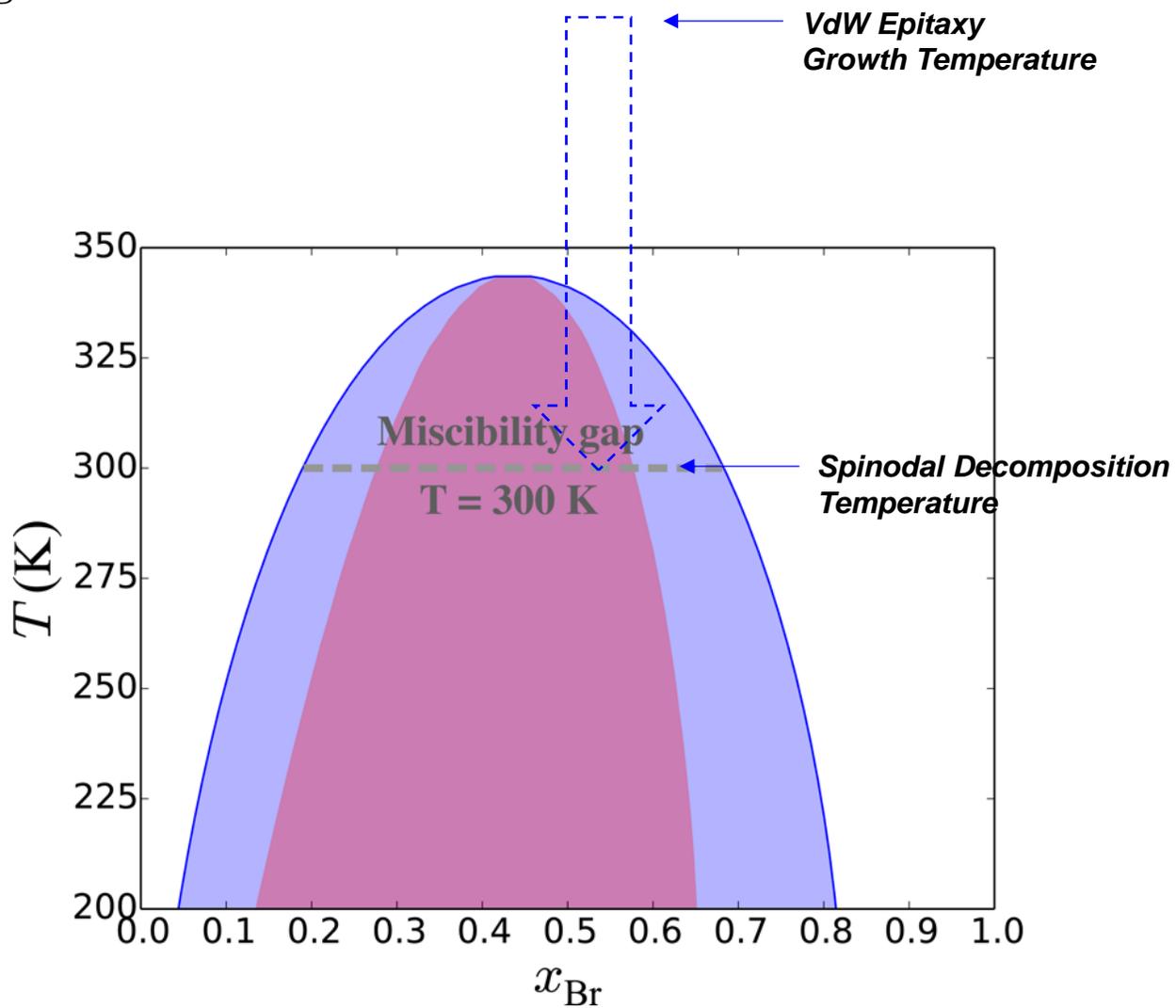

**Fig. S2** Adapted phase diagram of I/Br perovskite predicted by A. Walsh et al[5]. The simulation results indicate spinodal decomposition would occur within the purple region while binodal decomposition occurs in the blue region.

**Fig. S3**

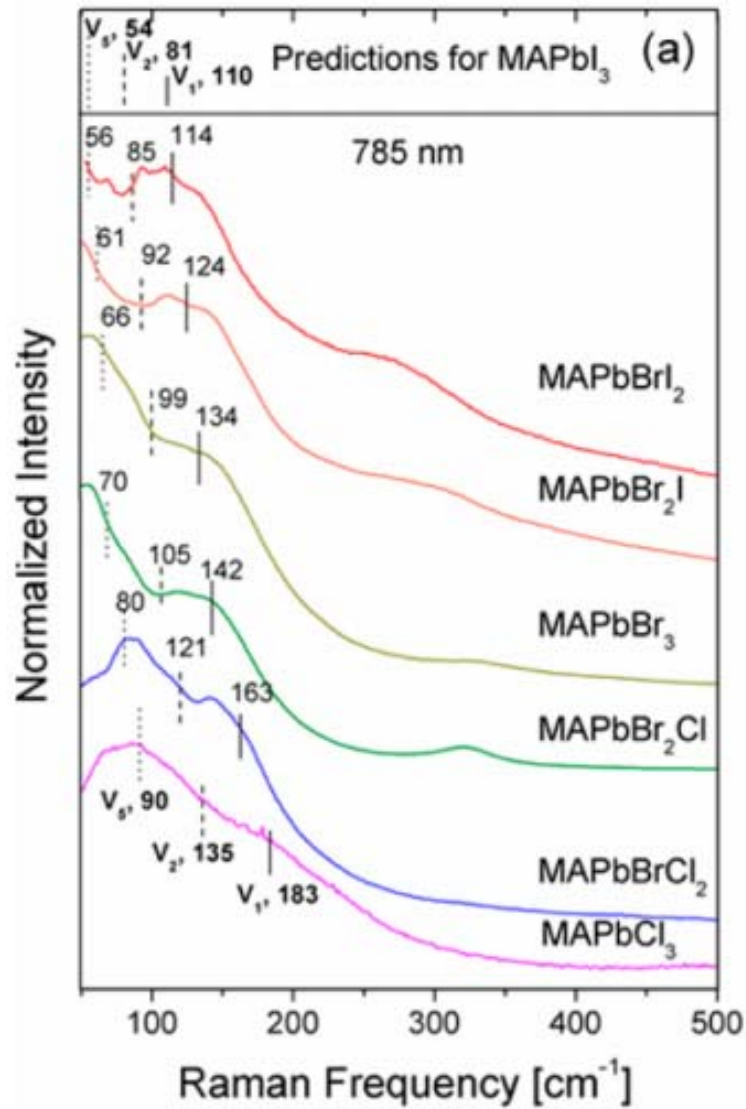

Fig. S3 Raman spectra for halide perovskites from P. Cameron et al[6]. The results indicate largely different LO phonon frequencies with different halide contents. Such discrepancy (e.g. in Pb-X stretching mode, X indicates halide element) would result in phonon confinement when a double heterojunction is formed. This is very common in III-V system (e.g. in AlAs/InAs)[7].

**Fig. S4**

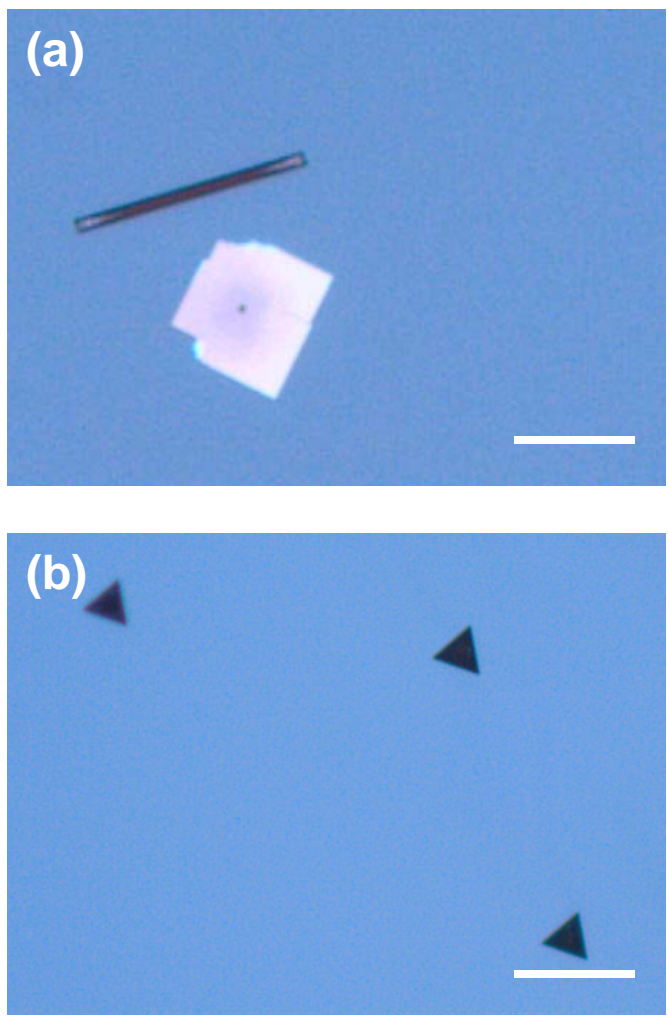

Fig. S4 Optical images of the single phase mixed halide perovskite. (a) thick rod and square sheet. (b) bulk pyramid-shaped crystallite. The morphologies were obtained at higher temperature zone where Wulff ripening is dominat. (scale bar: (a) 25 μm; (b) 30 μm; )

**Fig. S5**

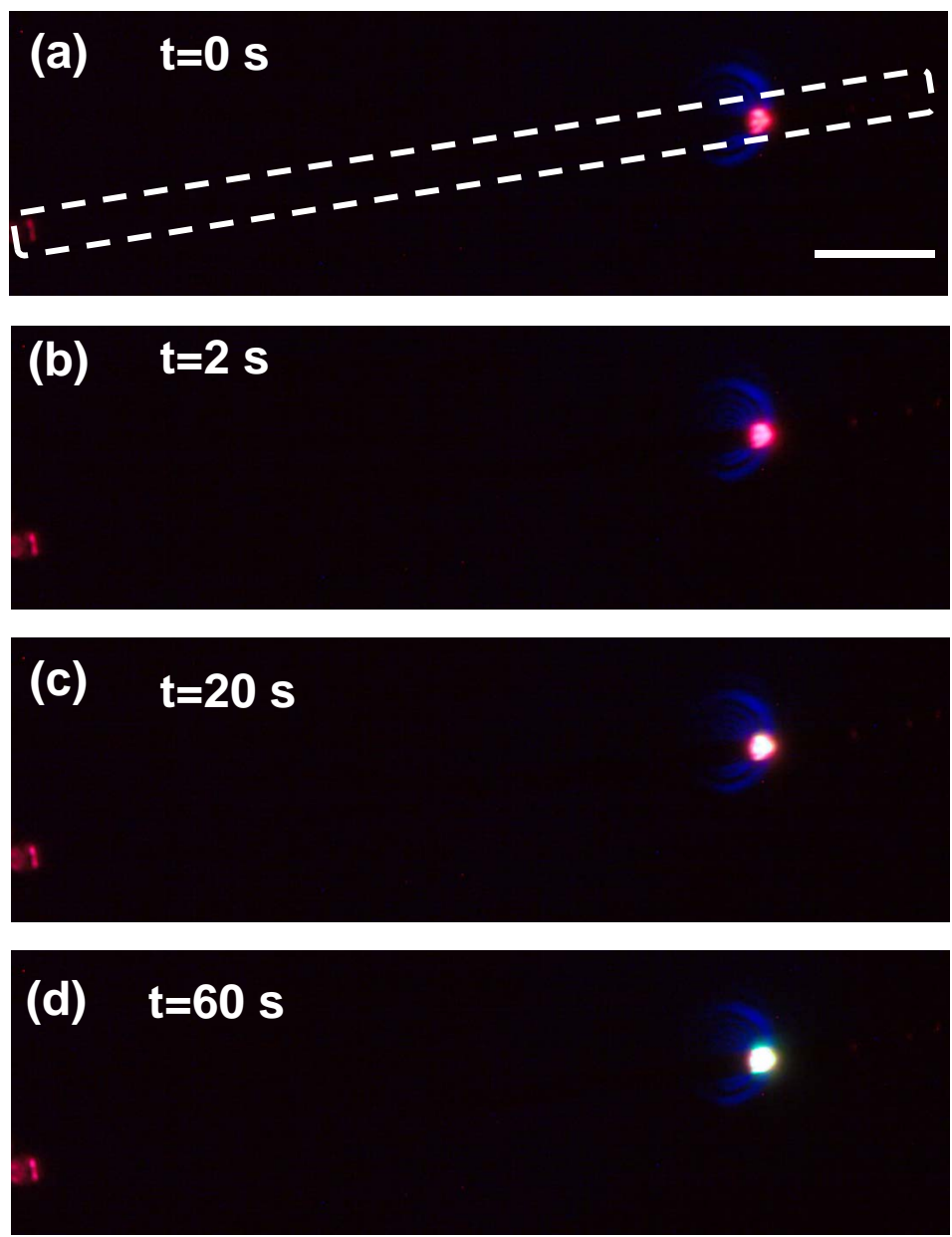

**Fig. S5 Time lapsed PL images of a mixed halide perovskite rod upon laser illumination. A similar phenomenon can be observed as described in the Fig. 2 of the main text. A stage of "dim PL" occurs at t = 0 s but rapidly becomes much brighter starting from t = 2 s. Green PL from the Br-rich phase can be seen at t = 60 s. Waveguide effect to the left of the images is also visible. The outline of the rod is highlighted by the dashed white line. (scale bar: 15 μm)**

**Fig. S6**

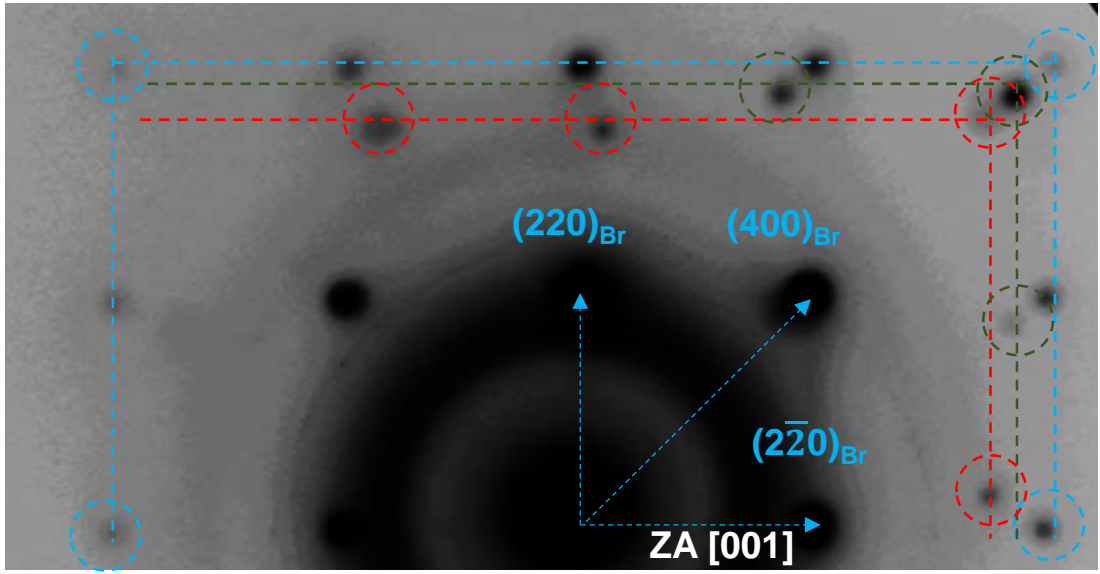

**Fig. S6 Reprint of the SAED pattern at the interface of the heterojunction in Fig. 2(l) after adjusted brightness/contrast.**

**Fig. S7**

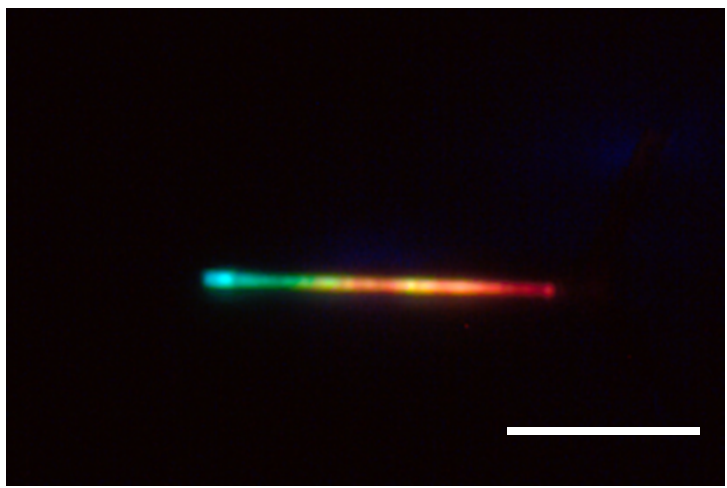

**Fig. S7** PL image of a mixed halide perovskite rod exposed to indoor light for two weeks after growth. The gradual change of PL color indicate a blend of different phases and supports our hypothesis of spinodal decomposition and coarsening. Room light is of much lower intensity and thus slows down the overall kinetics. (scale bar: 15 μm)

**Fig. S8**

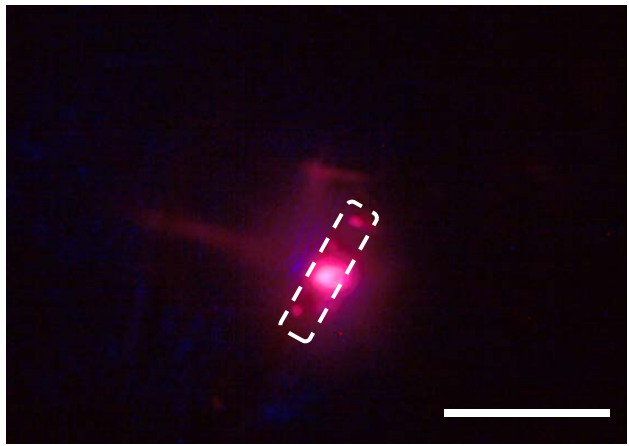

**Fig. S8 PL image of a mixed halide perovskite rod after annealing at 100 °C for 30 mins. The bright red PL color indicates that heating alone can not trigger significant phase separation or decompose the perovskite structure. This further demonstrates that the presence of light is essential for the phase separation to take place. (scale bar: 15 μm)**

**Fig. S9**

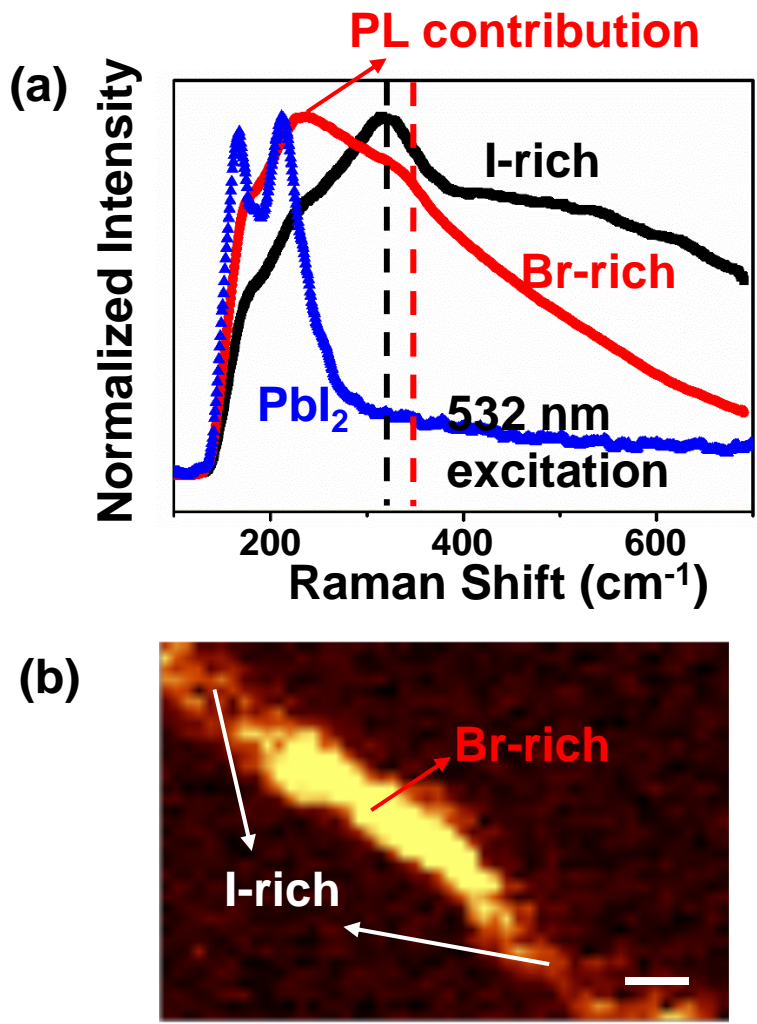

**Fig. S9** Raman characterization of the mixed perovskite halide double heterojunction. (a) Raman spectra of the Br-rich phase (red), I-rich phase (black) and pure single crystal $PbI_2$ flake (blue). The double heterojunction result showed a stark difference from the $PbI_2$, thus excluding the possibility of the degradation of the perovskite into $PbI_2$. The vibrational peaks corresponding to the torsion of MA ion[6] are different for the two phases as shown by the vertical dashed line. The Raman spectrum for the Br-rich phase is dominated by its PL since a 532 nm excitation is used. Based on this, a Raman mapping can be done to clearly reveal the two phases, as shown in (b), where the brighter part is the Br-rich phase. (scale bar: 1 μm)

# Fig. S10

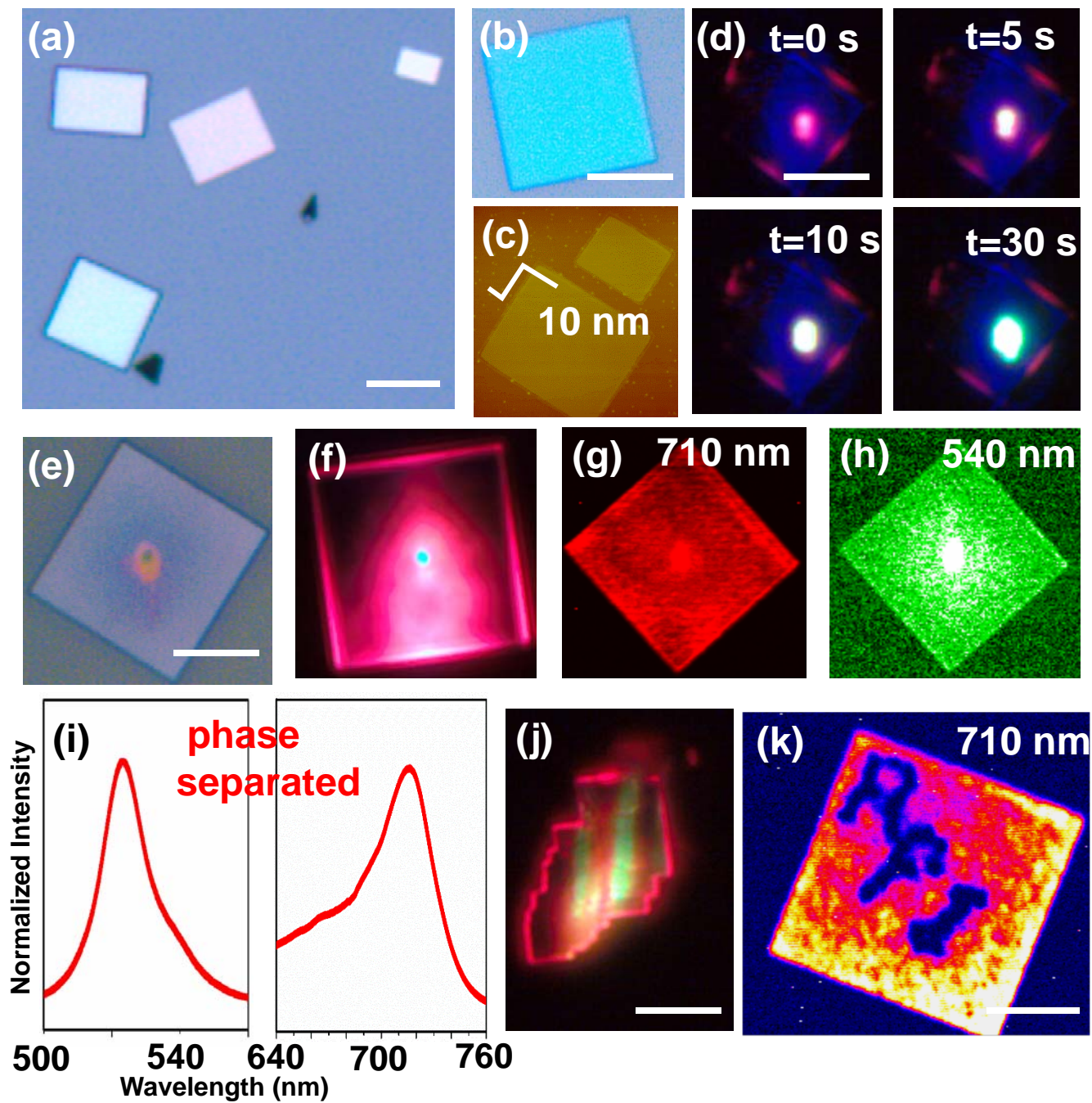

**Fig. S10** Growth, fabrication and characterization of the 2D mixed halide heterojunction. See the following page for detailed discussion.

Fig. S10 shows the 2D halide perovskite heterojunction based on mixed halide perovskite thin film. Shown in Fig. S10(a) and (b) are the optical images of as prepared $MAPbI_xBr_{3-x}$ square thin film on mica substrates, with the lateral size up to several tens of microns and thickness around 100 nm. With appropriate growth conditions, the pure Br-based perovskite could be as thin as 10 nm, as shown by the AFM image of Fig. S10(c). We have chosen relatively thicker film for the fabrication of 2D heterojunction out of the stability concern. Fig. S10(d) shows a time-dependent PL image change upon laser illumination similar to Fig. 2(e) – (h), with the only difference at the initial stage where we do not seem to observe the short-lived "dim PL". We attribute this phenomenon to the larger surface/bulk ratio of the thin film which may enable even room light to trigger the spinodal decomposition. Also notable from the PL image is that for the square thin film perovskite film, only red PL leakage at certain locations along the edge can be seen, which demonstrates the good crystallinity and intact film morphology before and after laser illumination. Fig. S10(e) shows a single square film after phase separation (the illuminated region located in the middle) and still the contrast of different phases is visible. With uniform illumination in Fig. S10(f), distinctive red and green PL is observed. The PL mapping in Fig. S10(g) – (h) further shows very similar PL intensity distribution with the 1D heterojunction, with the I-rich phase possessing larger intensity at the interface. Fig. S10(i) further shows the PL spectrum of the 2D heterojunction after phase separation. Compared with Fig. 3(b) we may see while the first peak at the green regime almost remain identical, the peak for the I-rich phase becomes narrower. With the aid of confocal microscope, we would be able to write patterns on the 2D thin film. To illustrate the idea, we present in Fig. S10(j) the PL image of Br-rich phase stripes obtained by shining uniform laser beam onto a photo-masked thin film and in Fig. S10(k) the PL mapping of a "RPI" pattern. Our demonstration shows that with adequate control of the light illumination, a 2D heterojunction with desired features could be achieved. (scale bar: (a-b, d-h) 10μm; (j) 8 μm; (k) 10 μm)

**Fig. S11**

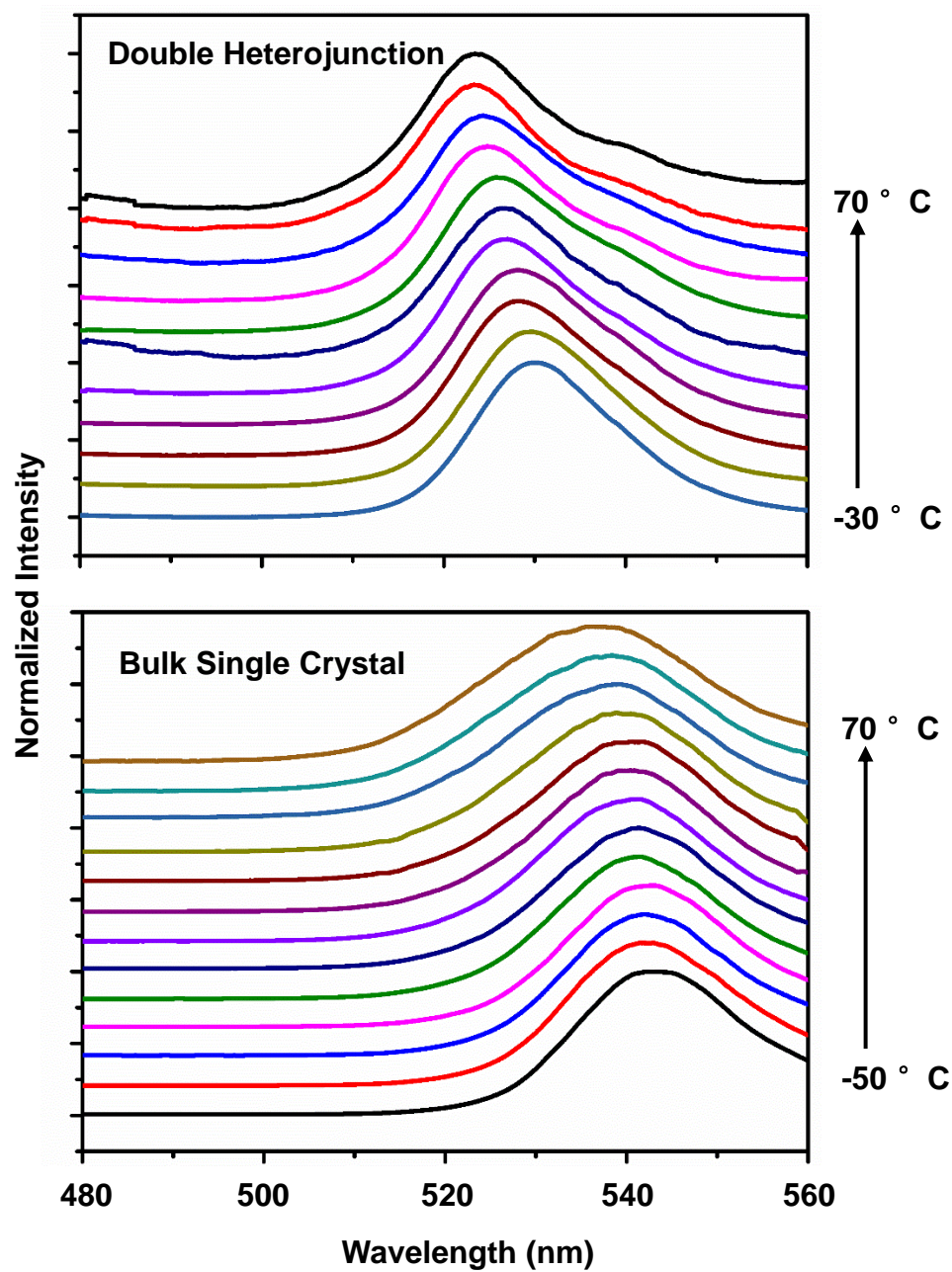

**Fig. S11** Temperature dependent PL spectra of the Br-rich phase in double heterojunction (upper) and the bulk single crystal (lower). While both types show a red shift of peak with decreasing temperature, the heterojunction displays a clear reduction of peak linewidth compared with that of the bulk single crystal.

## SI References